# Experimental generation of tripartite telecom photons via an atomic ensemble and a nonlinear waveguide

Dong-Sheng Ding[†,1,2], Wei Zhang[†,1,2], Shuai Shi[†,1,2], Zhi-Yuan Zhou[1,2], Yan Li[1,2], Bao-Sen Shi[1,2,*] and Guang-Can Guo[1,2]

[1]*Key Laboratory of Quantum Information, University of Science and Technology of China, Hefei, Anhui 230026, China and*

[2]*Synergetic Innovation Center of Quantum Information & Quantum Physics, University of Science and Technology of China, Hefei, Anhui 230026, China*

[†]*These authors contribute to this article equally*

*Corresponding author: \*drshi@ustc.edu.cn*

Non-classical multi-photon and number states attracts many people because of their wide applications in fundamental quantum mechanics tests, quantum metrology and quantum computation, therefore it is a longstanding aim to generate such states experimentally. Here, we prepare photon triplets by using the spontaneously Raman scattering process in a hot atomic ensemble cascaded by the spontaneous parametric down conversion process in a periodical poled nonlinear waveguide, the strong temporal correlations of these three photons are observed. Our experiment represents the first combination of the different order nonlinear processes and different physical systems, showing the feasibility of such composite system in this research direction. In addition, the all photons in the prepared genuine triplet are in telecom band make them be suitable for long-distance quantum communication in optical fibre.

Multipartite entanglement and correlations have huge potential applications in fundamental quantum mechanics and quantum information fields, therefore many physical systems have been used to generate successfully a triplet or to an eight-photon state [1-17], significant progresses have been made with the successful demonstrations of reducing quantum communication complexity [2, 10], topological error correction [11], and fundamentally testing non-locality [3, 12], etc, Besides, about 1000 entangled modes in continuous variables field has been also prepared

[17].

Many different techniques and physical systems are used for the direct generation of photon triplets. These include the tri-excitations in quantum dots [18], the combination of second-order nonlinear processes [19] and the high-energy electron-position collisions [20]. More recently, the technique of using cascaded photon pair sources prepared through nonlinear spontaneous parametric down conversion (SPDC) process in a nonlinear crystal is used for preparing genuine tripartite photons [14, 15]. Furthermore, a heralded two-photon photonic polarized entangled state without post-selection is realized [16] by this way. There is another way to generate tripartite photons through two SPDC processes cascaded by a sum-frequency process [21-23], where the single photon nonlinear interaction is demonstrated. There is one thing we want to pint out is that all protocols for preparing the triplets use the same physical systems, (for example two pp crystals in Refs [14-17], three pp crystals in Refs. [23]), and the same order nonlinear process, (for example the second-order in Refs. [14-17, 20, 23], and the third-order in Ref. [24]). In future, a quantum network must include many different physical systems and different interaction processes, therefore generating a photon triplet using the combination of different physical systems is interesting and deserved to be investigated.

Here, we realize the preparation of tripartite photons with the combination of two different physical systems for the first time: a hot atomic ensemble and a nonlinear waveguide. In our work, the spontaneously Raman scattering (SRS) process in a hot Rb atomic cell in a collinear configuration is used to prepare the non-classical photon pair source at 780 nm and 1530 nm firstly. Due to the use of the collinear configuration, the qualities of the correlated photons source are significantly better than the previous sources by using a vapour cell [26-29] (has a higher signal-to-noise ratio (SNR) and higher brightness). Then, the photon at 780 nm is sent to a periodically poled lithium niobate (PPLN) waveguide as a pump photon for generating two non-classical photons at 1551 nm and 1571 nm via SPDC. Experimentally we obtain the triple coincidence with a SNR of 11.03, clearly showing the strong temporal correlation of the created triplets. This is the first reported preparation of photon triplets using a $\chi^{(3)}$ nonlinear interaction cascaded by a $\chi^{(2)}$ process. Besides, the correlated three photons have a wavelength matched for optimal transmission in optical fibres, therefore all are suitable for long-distance quantum communications. Our source also allows experimental preparation of the tripartite entanglement without post-selection which is suitable for linear optical quantum computing.

**Experimental Results**

We use Rubidium 85 atoms to generate two non-classical correlated photons at the wavelengths of 780 nm and 1530 nm respectively, the layout of the experiment is depicted by Fig. 1. We use two lasers at wavelengths of 1475 nm and 795 nm, both from external-cavity diode lasers (DL100, Toptica), acting as pump 1 and pump 2, to generate the non-classical correlated photons. Both lasers have 1 MHz bandwidth. The pump 1 laser is red-detuned 0.8 GHz with the atomic transition of $5S_{1/2}(F=2)\text{->}5P_{1/2}(F'=2)$. The $^{85}$Rb atomic cell is heated to be 86 ℃, corresponding to the atomic density of $2.7 \times 10^{12} \text{cm}^{-3}$. Energy levels of |1>, |2>, |3> and |4> consist of a diamond atomic transition configuration. Due to the conservation of energy and momentum in nonlinear SRS process, the generated signal 1 and signal 2 photons have a strong correlation in time domain. Then the signal 2 photons are sent to a nonlinear PPLN waveguide and subsequently down-converted to two non-classical correlated signal 3 and signal 4 photons at wavelengths of 1551 nm and 1571 nm via SPDC via SPDC probabilistically. Thus, the signal 1, signal 3 and signal 4 photons are tripartite correlated photons.

The whole process could be described by a standard quantum optics model, the interaction Hamiltonian for SRS process can be written as the form of $\tilde{H}_1 = \gamma_1 \alpha \beta (a_{s1}^\dagger a_{s2}^\dagger + h.c)$. ($h.c$ is the Hermitian conjugate) where, $\gamma_1$ describes the coupling strength of the $\chi^{(3)}$ interaction. $\alpha, \beta$ is the electromagnetically field amplitude of pump 1 and pump 2 lasers respectively. $a_{s1}^\dagger, a_{s2}^\dagger$ denotes the creation operator of signal 1 and signal 2 photons. The nonlinear SPDC process could be represented by $\tilde{H}_2 = \gamma_2 (a_{s2} a_{s3}^\dagger a_{s3}^\dagger + h.c)$, where $\gamma_2$ describes the coupling strength of the $\chi^{(2)}$ interaction; $a_{s3}^\dagger, a_{s4}^\dagger$ denotes the creation operator of signal 3 and signal 4 photons. This whole process could be written as: $\exp(-i\tilde{H}_2)\exp(-i\tilde{H}_1)$. After the cascaded processes, the final state is:

$$\exp(-i\tilde{H}_2)\exp(-i\tilde{H}_1)|0_{s1}0_{s2}0_{s3}0_{s4}\rangle = -\gamma_1 \gamma_2 \alpha \beta |1_{s1}0_{s2}1_{s3}1_{s4}\rangle, \quad (1)$$

The signals 1, 3 and 4 are in triplet state.

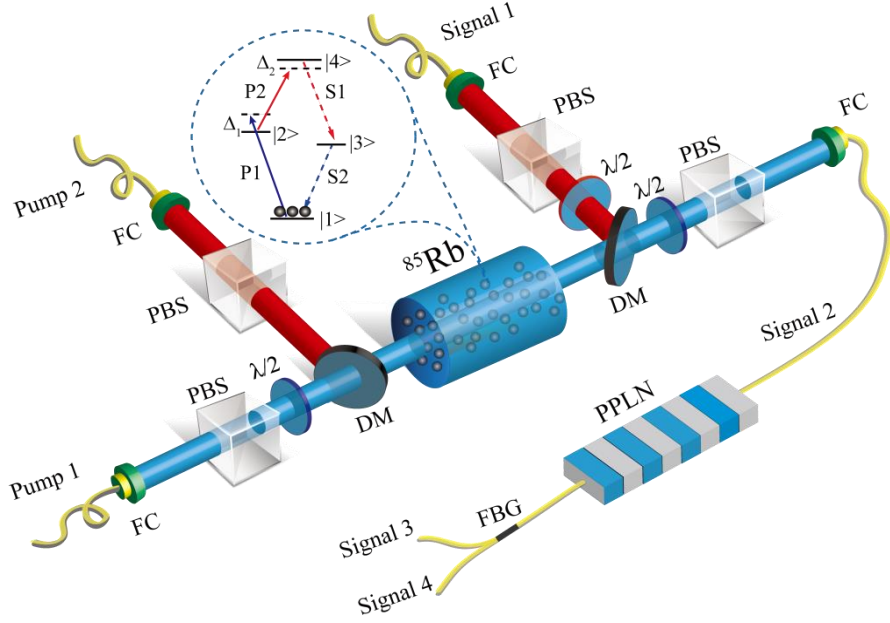

Fig. 1 The simplified experimental setup for generating tripartite photons. The picture in dashed circle is the energy level diagram for SRS process. λ/2: half-wave plate. DM: dichroic mirror, by which the light in telecom band is reflected and the visible light can transmit through it. FC: fibre coupler. PBS: polarization beam splitter. FBG: fibre Bragg grating (reflecting 1475 nm laser and transparent for 1530 nm light). P1: Pump 1. P2: Pump 2. S1: Signal 1. S2: signal 2.

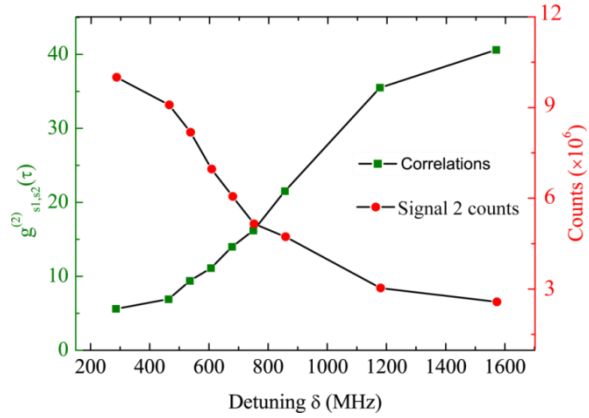

Fig. 2 The green line is the measured cross-correlation between signal 1 and signal 2 photons against the two-photon detuning δ, which is defined as $\Delta_2$-$\Delta_1$. The red line is the single count of signal 2 photons against detuning δ, including the losses of collection and detection.

We first prepare a very bright non-classical correlated photon pair source using the collinear SRS in a hot Rb atomic vapour cell. The pumps 1 and 2 propagate collinearly through the cell with the aid of a dichroic mirror, their powers are 115 mW and 12.5 mW, the beam waists are 60 μm and 118 μm respectively. Due to momentum conversation in SRS process, the correlated photons signal 1 and signal 2 at 1530 nm and 780 nm are emitted in same direction. We use three interference filters (Semrock LL01-780 with bandwidth of 2 nm, 98% transmittance) to reduce the noise level of pump 1 below $10^{-18}$, and use another three interference filters (SemrockNIR01-1530 with bandwidth of 3 nm, 98% transmittance), a longpass filter (Thorlab, FELH 1000) and a fibre Bragg grating to reduce the noise from pump 2 laser beam below the order of $10^{-20}$. The emitted signal 1 and signal 2 photons are collected by single-mode fibres with a coupling efficiency of 50% and 70% respectively and then are detect by two detectors D1 (avalanche diode, PerkinElmer SPCM-AQR-15-FC) and D2 (a free running In-GaAs Photon Detector, ID QuantiqueID220-FR-SMF). The signals from both detectors are sent to a time-correlated single photon counting system (TimeHarp 260) to measure the time-correlated function. The heralded efficiency between signal 1 and signal 2 photons is 3%.

We change the two-photon detuning by adjusting the frequency of pump 2 laser, then record the single count of signal 2 photons and the cross-correlation rate between signal 1 and signal 2 photons. The results are given in Fig. 2. When the two photon detuning is 1570 MHz, the single count of the generated signal 2 photons is $1\times10^6$ per second and the violation $R$ is $1.37\times10^4$ with the power of pump 1 laser of 50mW, (see supplements). This photon source gives a platform with a competitive photon counts with the non-classical photon source using SPDC in a nonlinear waveguide. In the experiment for preparing tripartite photons, the detuning is set to be 850 MHz, the obtained signal 2 photon count is $6\times10^6$ per second in this case.

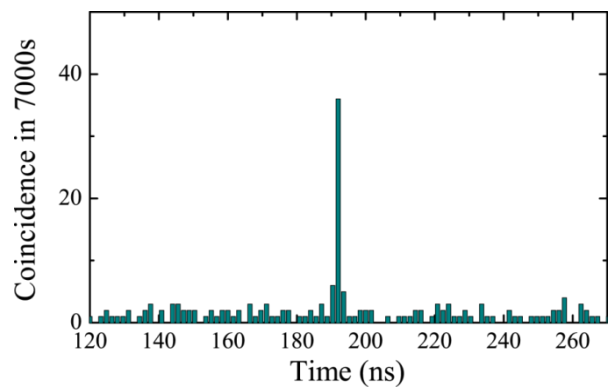

Fig. 3.The measured coincidence rate between signal 3 and signal 4 when $6\times10^6$ per second signal

2 photons are input to PPLN waveguide. In this process, two free running detectors are used to measure the coincidence counts. The SNR is 16.7.

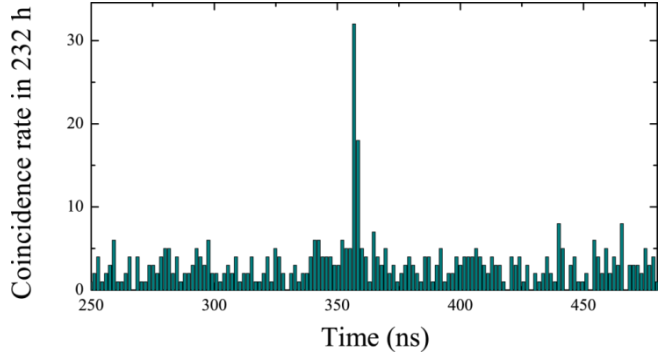

Fig. 4. The three photons coincidence rate measured in 232 hours. The maximum tripod event is 32. The noises are about 2.9 counts.

Next, we input the prepared signal 2 photons into a 5-cm long PPLN waveguide (with a conversion efficiency of $10^{-6}$) for acting as a pump light for further SPDC experiment. The generated photons of signal 3 and signal 4 around 1551 nm and 1571 nm are distinguished by a fibre Bragg grating with each output of 10-nm bandwidth. We measure the correlation between the photons by using two free running In-GaAs Photon Detectors. The dark counts of these detectors are both set to be 400/s, the dead times are set to be 1μs. The experimental results are shown in Fig. 3, the measured time is 7000s. The coincidence rate between signal 3 and signal 4 photons is about 18.5/h, the heralded efficiency is 3%.The measured maximum cross-correlation rate is 16.7.

At last, we focus on the most important part in our experiment: confirming the correlations of tripartite photons. The signal 3 photons are detected by D3 (In-GaAs Photon Detector, the Princeton Lightwave Single Photon Benchtop Receiver PGA-604), the signal 4 photons are detected by a free running In-GaAs Photon Detector D4. The detector D3 works in gate mode, which is triggered by the electronic pulse from detector D4, the electronic pulse is delayed by a delay generator (DG535). We adjust the delay of DG535 to make signal 3 and signal 4 photons be in coincidence window, then measure the coincidence rate between detector D2 and the output signal from detector D3. The result is shown in Fig. 4, the total measurement time is 232 hours.

We obtain a clear sharp triple coincidence peak with the SNR of 11.03, the maximum correlated tripod count is 32, clearly showing the strong temporal correlation of the created triplets. The overall triplets in 232 hours are measured as 50. The theoretical estimation of triplet's production rate is about $1.925 \times 10^4$/h (5.3/s) with the considerations of all losses from the coupling and detection (the efficiency of all detectors is 10%). This generation rate is better than the rate in Ref. [15] in which 45 triplets/min (0.75/s) are generated. The recorded triplets could be significantly increased if a higher efficient detector like a superconductor detector used in Ref. 15 is available in our experiment.

In order to further check the non-classical correlation properties of three photons, we use an inequality $|u \cdot v \cdot \kappa| \leq |u||v||\kappa|$ defined in supplements to construct an extended Cauchy-Schwarz inequality for three-photon coincidence,

$$R_3 = \frac{[g^{(3)}_{s1,s3,s4}(\tau_1,\tau_2)]^2}{g^{(2)}_{s1,s1}(0) g^{(2)}_{s3,s3}(0) g^{(2)}_{s4,s4}(0)} \leq 1 \qquad (2)$$

where, $g^{(3)}_{s1,s3,s4}(\tau_1,\tau_2)$ is the correlation of signal 1, signal 3 and signal 4 photons, $\tau_1$ is the relative delay time of signal 3 and signal 4, and $\tau_2$ is the relative delay time between signal 1 and signal 3. $g^{(2)}_{s1,s1}(0)$, $g^{(2)}_{s3,s3}(0)$, $g^{(2)}_{s4,s4}(0)$ is the auto-correlation of signal 1, signal 3 and signal 4 photons. In our experiment, $g^{(3)}_{s1,s3,s4}(\tau_1,\tau_2) = 11.03$, $g^{(2)}_{s1,s1}(0) = 1.16$. According to the measured auto-correlation of signal 3 and signal 4 generated with an input weak coherent 780 nm laser beam, we estimate the auto-correlations of signal 3 and signal 4 are $g^{(2)}_{s3,s3}(0) = 1 \pm 0.05$, $g^{(2)}_{s4,s4}(0) = 1 \pm 0.05$ respectively (see supplements). The calculated value $R_3 = 105.67 \pm 10.54$, is much large than 1, therefore the inequality is strongly violated. Thus, there is a strong non-classical correlation in these three photons. In our coincidence system, the time $\tau_1$ is fixed in coincidence window of signal 3 and signal 4 by using a delay generator DG535. $\tau_2$ is 356.8 ns, measured according to experimental data from Fig. 4.

**Discussions**

Our experiment shows some significant differences comparing with the previous experiments for generating genuine tripartite photons:1). In our scheme, the input pump photon for the second nonlinear process is generated using SRS in an atomic vapor cell. We use a collinear configuration to increase the nonlinear interaction length to improve the photon number, such a configuration is never considered in a hot vapor cell before due to the bad SNR. We have overcome the problem of filtering, a high SNR is achieved, therefore we use this configuration. 2) The tripartite photons are all in telecom band, therefore all are suitable for long-distance quantum communications. We can also prepare a GHZ state with all photons in telecom band without any post-selection.

**Conclusions**

In this letter, the cascaded two different order nonlinear processes are used for generating tripartite photons, the experimental data clearly show the strong temporal correlation among these photons. The cascade scheme in our experiment includes two different order nonlinear processes: $\chi^{(3)}$ and $\chi^{(2)}$ nonlinear interactions, and two different physical systems: an atomic ensemble and a nonlinear waveguide. In addition, all three photons have a wavelength matched for optimal transmission in optical fibres, therefore they are suitable for long-distance quantum communications. Our experiment opens a pave forward the quantum link between different systems by different nonlinear interactions in future.

**Author contributions**

**Acknowledgements**

We thank Dr. Yun-Feng Huang, Dr. Fang-Wen Sun and Dr. Yun-Kun Jiang for helpful discussions. This work was supported by the National Fundamental Research Program of China (Grant No. 2 011CBA00200), the National Natural Science Foundation of China (Grant Nos. 11174271, 61275115，61435011), the Youth Innovation Fund from USTC (Grant No. ZC

# Supplementary Material for

# Experimental generation of tripartite telecom photons via an atomic ensemble and a nonlinear waveguide

Dong-Sheng Ding, Wei Zhang, Shuai Shi, Zhi-Yuan Zhou, Yan Li, Bao-Sen Shi, and Guang-Can Guo

**Preparation of signal 1 and signal 2 photons from a hot atomic vapors cell.**

As stated in the main text, we must perform and improve the efficiency of the preparation of non-classical correlated photon pair from an atomic vapor cell. The cross-correlation and auto-correlation between the generated photons are measured, which are shown in figure S1. The two-photon detuning is 1570 MHz. The power of pump 1 laser is about 50mW. The single count of signal 2 photon is $1 \times 10^6$ per second, which is comparable with the non-classical photon source by SPDC in a nonlinear waveguide. The experimentally measured cross-correlation and auto-correlation are:

$$g^{(2)}_{s1,s2}(\tau = 26ns) = 126.7 \,;\, g^{(2)}_{s2,s2}(0) = 1.17 \,;\, g^{(2)}_{s1,s1}(0) = 1.16. \quad (1)$$

The non-classical correlation between the generated photons can be proved by checking whether the Cauchy-Schwarz inequality is violated. Usually classical lights satisfy the following equation [1]:

$$R = \frac{[g^{(2)}_{s1,s2}(\tau)]^2}{g^{(2)}_{s1,s1}(0) g^{(2)}_{s2,s2}(0)} \leq 1 \quad (2)$$

where, $g^{(2)}_{s1,s2}(\tau)$, $g^{(2)}_{s1,s1}(0)$ and $g^{(2)}_{s2,s2}(0)$ are the cross-correlation and auto-correlations of the photons respectively, the time $\tau$ is the correlated time between signal 1 and signal 2 photons. According to the experimental results in figure S1, the calculated $R = 1.37 \times 10^4$.

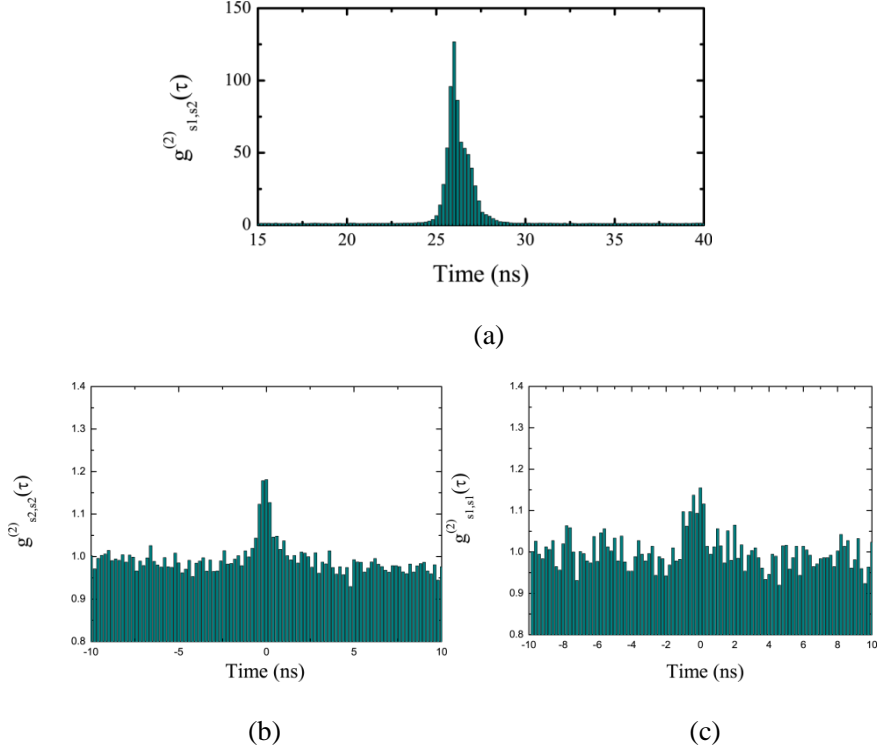

Fig. S1. (a) The cross-correlation rate between signal 1 and signal 2 photons. (b) The auto-correlation of signal 1 photons. (c) The auto-correlation of signal 2 photons

**Coincidence window Shift with detuning of pump 1 laser.**

The laser with wavelength of 795 nm is stabilized with the spectrum locking technique. It is red-detuned 1.45GHz with the $^{87}Rb$ atomic transition of $5S_{1/2}(F=1)$->$5P_{1/2}(F'=1)$ by using a 1.5 GHz acousto-optic modulator. In our experiment, if we change the detuning of 795 nm laser, the window of time-correlation would shift because of the delay of 780 nm photon. The reason is that the linear susceptibility of the generated 780 nm photon is modulated by the detuning of 795 nm laser, please refer to Ref. 2 for the detail of this FWM delay effect. So we need to stabilize the 795 laser over long period experimental performance.

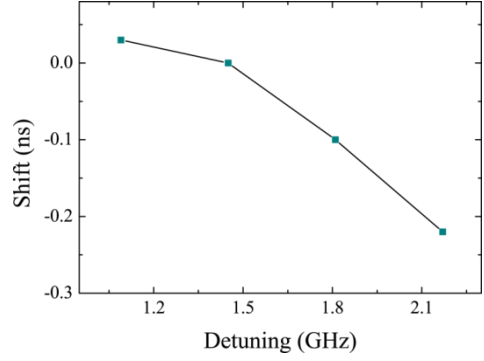

Fig. S2. The coincidence window is shifted by modulating the detuning of pump 1 laser.

**Bandwidth of photon pair against the temperature of cell.**

Due to the Doppler-effect, the bandwidth of the generated photon is broaden. For different temperature of the cell, the bandwidth of generated photon pair is different. We measure the bandwidth of the generated photons against the temperature of cell and obtain the relationship between the bandwidth of photon and the temperature. With the increment of temperature in atomic cell, the bandwidth of generated photon becomes narrow. The results are shown in Fig. S3. In our experiment, the temperature is set to be 86 °.

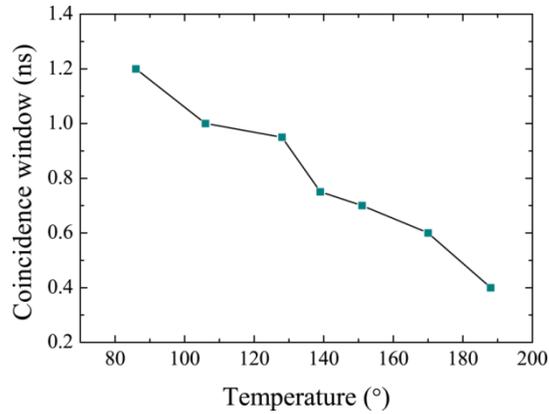

Fig. S3. The measured coincidence window of signal 1 and signal 2 photons against temperature of the cell.

**Proof of inequality for three photons coincidence.**

We assume $\mu$, $\nu$, $\kappa$ be the arbitrary random variables. According to Cauchy-Schwarz inequality for two variables $|v \cdot \kappa| \leq |v||\kappa|$, we construct a variable $|u|v \cdot \kappa$, and obtain:

$$|u||v \cdot \kappa| \leq |u||v||\kappa| \tag{3}$$

By using Cauchy-Schwarz inequality for two variables $u$ and $v \cdot \kappa$,

$$|u \cdot v \cdot \kappa| \leq |u||v \cdot \kappa| \tag{4}$$

Thus, we obtain the inequality for three variables $\mu$, $v$, $\kappa$:

$$|u \cdot v \cdot \kappa| \leq |u||v||\kappa| \tag{5}$$

**Estimation of auto-correlation of signal 3 and signal 4 photons.**

We input a weak coherent laser at 780 nm into the PPLN waveguide to generate and then measure the auto-correlation of signal 3 and signal 4 photons against different power of 780 nm laser. The measured result is shown in figure S4. With a decrement of power of 780 nm laser, the probability of multi-photons of signal 3 and signal 4 would decrease. By using this process, the auto-correlation of signal 3 and signal 4 is estimated to be $g^{(2)}_{s3,s3}(0) = g^{(2)}_{s4,s4}(0) = 1 \pm 0.05$

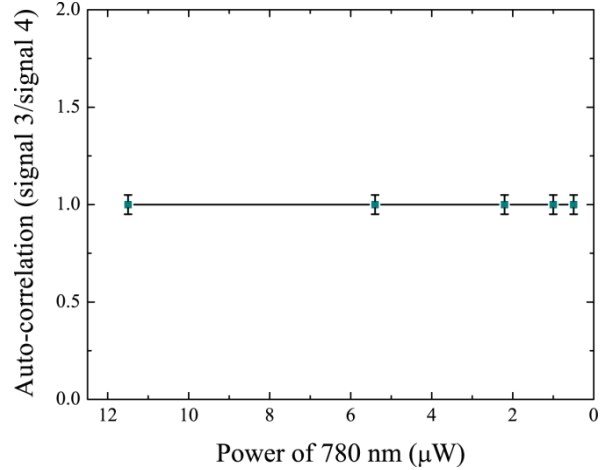

Fig. S4. The measured auto-correlation of signal 3 and signal 4 photons against the power of 780 nm laser. Error bars are the experimental statistic.